# A framework for reuse of multi-view UML artifacts

Hamza Onoruoiza Salami*, Moataz Ahmed

Information and Computer Science Department,
King Fahd University of Petroleum and Minerals
Dhahran 31261, Saudi Arabia
e-mail: {hosalami, moataz}@kfupm.edu.sa

*Abstract*— **Software is typically modeled from different viewpoints such as structural view, behavioral view and functional view. Few existing works can be considered as applying multi-view retrieval approaches. A number of important issues regarding mapping of entities during multi-view retrieval of UML models is identified in this study. In response, we describe a framework for reusing UML artifacts, and discuss how our retrieval approach tackles the identified issues.**

## I. INTRODUCTION

Software reuse refers to the development of software using previously developed software, rather than from scratch [1]. There are several benefits of software reuse such as accelerated development, reduced overall cost, reduced risk and effective use of specialists [2]. However, the drawbacks of software reuse include increased maintenance cost, lack of tool support, effort to find and adapt reusable components, effort to create and maintain components, and *not-invented-here* syndrome [2, 3].

Software reuse is not restricted to source code reuse. Other artifacts like domain models, requirement specifications, design, documentation and test data can be reused as well [4]. The benefits of reuse can be maximized if early-stage artifacts are reused, because it leads to reuse of later-stage artifacts derived from the early-stage artifacts [5].

There are four phases of software reuse; representation, retrieval, adaptation and incorporation [6]. At the representation stage, a model of the new software component (query) is presented. During retrieval, a software component which is similar to the query, and whose adaptation cost is minimal is selected from the components library or repository. The retrieved component is modified to obtain a new component during adaptation. Finally, the new component is incorporated or integrated into the repository. Unified Modeling Language (UML) is a general-purpose modeling language maintained by the Object Management Group (OMG), a consortium of companies. It provides diagrams for visualizing, specifying constructing and documenting the artifacts of a software-intensive system [7]. UML is widely used during initial stages of software development such as requirements engineering, architectural and detailed design.

Significant amount of research has been carried out regarding reuse of early-stage artifacts represented using UML. Early-stage artifacts are usually modeled and analyzed from different perspectives such as structural and behavioral views. Yet, a review of current literature suggests that little research effort has been put in the development of techniques for reusing software artifacts described from multiple viewpoints. Thus in this research, we describe a framework for reusing artifacts described from structural, behavioral and functional perspectives. Requirement specifications for new software are compared with requirement specifications of existing software systems contained in a repository. The corresponding artifacts (for example design, code and documentation) for the software system with the most similar requirements are returned for reuse, because it is expected that systems with similar requirements should have many other artifacts in common.

We identify a number of issues that are not considered in the few existing multi-view UML artifact reuse approaches, and explain how our proposed reuse framework tackles these issues. The rest of the paper is organized as follows: Section II describes related work. In Section III, we describe a framework for reusing UML artifacts. Section IV discusses important issues to be considered during multi-view retrieval. Our approach for retrieving multi-view UML artifacts is the subject of Section V. Finally, we conclude the paper and describe future work in Section VI.

## II. RELATED WORK

The UML taxonomy of diagrams partitions the various diagrams into two categories: structure diagrams and behavior diagrams [8]. Structure diagrams such as class, component, object, composite structure, deployment, package and profile diagrams document the static structure of system objects. On the other hand, behavior diagrams like activity, use case, state machine, sequence, communication, interaction overview and timing diagrams show the dynamic behavior of system objects. In this section, we discuss existing research on UML artifact reuse. Table I summarizes the different existing works. We consider an existing reuse work as multi-view, if it matches artifacts that have at least one structure diagram and one behavior diagram. Otherwise, the work is considered as utilizing a single-view approach.

In [9] authors utilized the WordNet lexicon and Case Based Reasoning (CBR) for retrieving UML models. Because the authors represented cases as class diagrams, their work involved retrieval of UML class diagrams. Robles et al. [10] have used domain and application ontologies for class diagram retrieval. Domain ontologies are used to measure the semantic similarity between classifier names, while application ontologies are needed to measure the semantic similarity between class diagram classifiers and relationship types.

In [11] a set of similarity metrics was used to measure class diagram similarity based on the semantic relatedness of class names, class attributes and class methods. Authors in [12] computed the structural similarity of class diagrams using an inexact graph matching technique. The similarity score of class diagrams was calculated from their adjacency matrix representation using a lookup table containing difference values for the different types of class diagram relationships.

The similarity between two sets of sequence diagrams are computed using two nested levels of genetic algorithm (GA) in [13]. At the upper level, sequence diagrams in one model are mapped to the sequence diagrams in the other model. At the lower level the similarity between two sequence diagrams is measured by mapping classes in both diagrams, and considering the number of matching and differing method calls. A CASE tool was developed in [14] to automatically retrieve sequence diagrams from a repository using a graph matching algorithm called SUBDUE. The authors indicated that the matching technique can also be applied to use case diagrams and class diagrams, thus we classify their approach as a multi-view reuse approach.

Kotb [15] has described an approach for retrieving similar use case descriptions using textual entailment, a natural language processing technique. A text T entails another text H if the meaning of T can be inferred from H. Any repository use case whose summary of descriptions is entailed by that of the query is retrieved and automatically adapted for reuse. In [16] information extracted from use case diagrams are stored in an Ontology Web Language (OWL) base ontology. The ontology is stored in a relational database system which is queried during reuse.

Information retrieval techniques were used for scenario management and reuse in [17]. Use case scenarios were represented by a set of attributes such as goals, authors, events, actors and episodes. The similarity between two scenarios was computed as the degree of overlap between their attributes. Similarly, in [18] use cases were matched by computing a similarity measure of their event flow vectors.

In [19] query and repository UML models are transformed from their XMI representation to first order logic specifications. The specifications are then matched, guided by some set of rules. Their approach is deemed to be multi-view, because it supports matching of class, sequence, use case and communication diagrams.

In the RedSeeDS project [20], repository software systems are considered for reuse if their requirements are similar to those of the new system. Requirements were represented in Requirements Specification Language (RSL) [21] in three possible formats: scenarios written in less formal natural language sentences; scenarios written in more formal constrained Structured English sentences; and using UML sequence and activity diagrams. None of the UML structure diagrams are considered during retrieval; hence, we do not consider their work as a multi-view reuse work.

In [22] software models are retrieved for reuse in two steps: classification and retrieval. During classification, a model is described from six facets which capture its functional requirements and useful properties. Predefined terms for each facet are arranged on a conceptual graph to aid the retrieval process. In the retrieval stage, similarity between query and repository models are computed using either the shortest distance in the conceptual graph or the degree of overlap of descriptor terms for both models. Because software models containing class, object, activity, state machine and collaboration diagrams can be retrieved for reuse, we consider their work as utilizing a multi-view reuse approach.

Park and Bae [23] adopt a two-stage multi-view approach for retrieving repository artifacts. In the first stage, query and repository class diagrams' structures are compared using analogy. Based on the similarity scores, a subset of repository UML models are selected. During the second stage, the authors compute the similarity scores of graphical representations of sequence diagrams in the shortlisted models.

## III. PROPOSED REUSE SYSTEM

This section describes our proposed system for reusing software modeled using UML. As shown in Fig. 1, reuse is carried out in four steps: pre-filtering, multi-view retrieval, adaptation, and integration.

### A. Pre-filtering

The aim of the pre-filtering stage is to minimize retrieval time by selecting a first set of repository artifacts, which will be assessed and ranked in the following stage. Pre-filtering is particularly important when the repository contains many models, because it eliminates the need to load requirement specifications of *all* systems from the repository into the primary memory of the computer during retrieval. In this stage, metadata of the new requirement is compared with the metadata of each software system held in the repository. In order to ensure that this stage is computationally inexpensive, we propose using two classes of previously obtained metadata: *computed metadata and extracted metadata*. Both types of metadata are automatically obtained from requirement specifications when new software systems are stored to the repository for the first time, and whenever changes are made to repository models.

Previously *computed metadata* refer to easily computable size metrics such as total number of classes in a class diagram, number of messages exchanged by objects in a sequence diagram, number of attributes and operations of classes, and cyclomatic complexity of state machine diagrams. This set of metrics can be used to filter out repository requirement specifications whose sizes differ significantly from that of the new system.

Furthermore, the domain of the software system can be inferred from *extracted metadata* such as class and package names in the model. This provides a way of selecting only artifacts belonging to similar domains as the new system [9].

### B. Multi-view Retrieval

In this stage, matching and similarity metrics are employed to asses and rank the requirement specifications shortlisted in the pre-filtering stage. Matching with GA is explained in Section V. At the end of this stage, a ranked list of requirement specifications is presented to the reuser. Requirement specifications at the top of the list are most similar to the new requirement specifications, thus reuse of the corresponding artifacts (for example design, code and documentation) from the repository should require the least time and effort.

### C. Adaptation

During adaptation the reuser modifies the artifacts corresponding to the most similar requirement specifications to suit the needs of the new system.

### D. Integration

In this stage, all artifacts for the new system, as well as automatically computed or extracted metadata are stored in the repository. In order to avoid degradation of response time and memory requirements, only new software systems that are sufficiently different from existing software systems should be added to the repository.

### IV. IMPORTANT ISSUES FOR CONSIDERATION DURING MULTI-VIEW RETRIEVAL

In this section, we identify three important issues that should be taken into consideration during comparison of multi-view requirement specifications.

### A. Issue 1: Consistent Mapping of Classes in Class Diagrams and Sequence Diagrams

We provide an illustrative example to underscore this issue. Assume a requirement specification Q is to be compared with two requirement specifications R1 and R2 from the repository. Q, R1 and R2 each have one class diagram and one sequence diagram as shown in Fig. 2. While comparing requirement specifications Q and R1, a multi-stage retrieval technique might produce maximum similarity value for both classes in the first stage, and erroneously assign maximum similarity value for the sequence diagrams in the second stage (and vice versa depending on which set of diagrams are first matched). Similarly, a retrieval technique which merely computes multi-view similarity as weighted sum of single-view similarity values would produce a wrong aggregate similarity value. Both approaches produce inaccurate similarity scores because of the inconsistent mapping of classes in the class diagrams and sequence diagrams (A1:A2, B1:B2, C1:C2 and B1:C2, C1:B2 in the class diagrams and sequence diagrams, respectively).

In order to further appreciate why the similarity score between Q and R1 should not be maximum, consider requirement specifications Q and R2. From Fig. 2e, class C3 is *composed of* class D3. If both classes are merged into one class, R2 will become identical with Q. From the perspective of the reuser, merging the two classes (C3 and D3 in R2) may require less effort than resolving the inconsistency between Q and R1. Thus intuitively, we expect the similarity score between Q and R2 to be higher than that between Q and R1. We note that none of the previous multi-view reuse works identified in Section II (that is, [14, 19, 22, 23]) have addressed the issue of consistent mapping of classes in class diagrams and sequence diagrams.

### B. Issue 2: Consistent Mapping of Classes in Class Diagrams and State Machine Diagrams

State machine diagrams are used to model the behavior of system elements such as objects (that is, class instances) [8]. They show how an object responds to events according to its current state, and how it enters into new states [24]. Just as classes should be consistently mapped in class diagrams and sequence diagrams, it is important to ensure that classes are consistently mapped when comparing two models containing class and state machine diagrams.

Only reference [22] discusses retrieval of software artifacts containing state machine diagrams. In their work, the authors compared diagrams of different types at a very abstract level. For example, a taxonomy of different UML diagrams was built for the 'design view' facet. Using the taxonomy, the similarity between two models was computed from the distance between the types of diagrams contained in the models. The authors did not discuss the consistent mapping of classes in class and state machine diagrams.

### C. Issue 3: Efficient Mapping of Multiple Sequence Diagrams in two Requirements

During the requirements phase of a software project, use cases are used to specify the functionality of a system. One or more sequence diagrams is then used to realize each use case [4]. Thus, it is common for requirement specifications to contain several sequence diagrams. An important issue is how to efficiently compare the sets of sequence diagrams in two requirement specifications. None of the existing works on multi-view reuse has addressed this issue.

TABLE I.     UML ARTIFACTS SUPPORTED BY VARIOUS STUDIES

| Work (First Author & Publication Year) | Structure Diagrams | | Behavior Diagrams | | | | | | Multi-view |
|---|---|---|---|---|---|---|---|---|---|
| | Class Diagram | Object Diagram | Sequence Diagram | Use Case Diagram | Activity Diagram | State Machine Diagram | Collaboration Diagram | Communication Diagram | |
| Ahmed 2006, [13] | | | √ | | | | | | |
| Ali 2003, [22] | √ | √ | √ | | √ | √ | √ | | √ |
| Bonilla-Morales 2012, [16] | | | | √ | | | | | |
| Alspaugh 1999, [17] | | | | √ | | | | | |
| Blok 1998, [18] | | | | √ | | | | | |
| Gomes 2002, [9] | √ | | | | | | | | |
| Khalifa 2008, [19] | √ | | √ | √ | | | | √ | √ |
| Kotb 2010, [15] | | | | √ | | | | | |
| Park 2010, [23] | √ | | √ | | | | | | √ |
| Robinson 2004, [14] | √ | | √ | √ | | | | | √ |
| Robles 2012, [10] | √ | | | | | | | | |
| Rufai 2003 [11] | √ | | | | | | | | |
| Salami 2012, [12] | √ | | | | | | | | |
| Bildhauer 2009 [20] | | | √ | √ | | | | | |

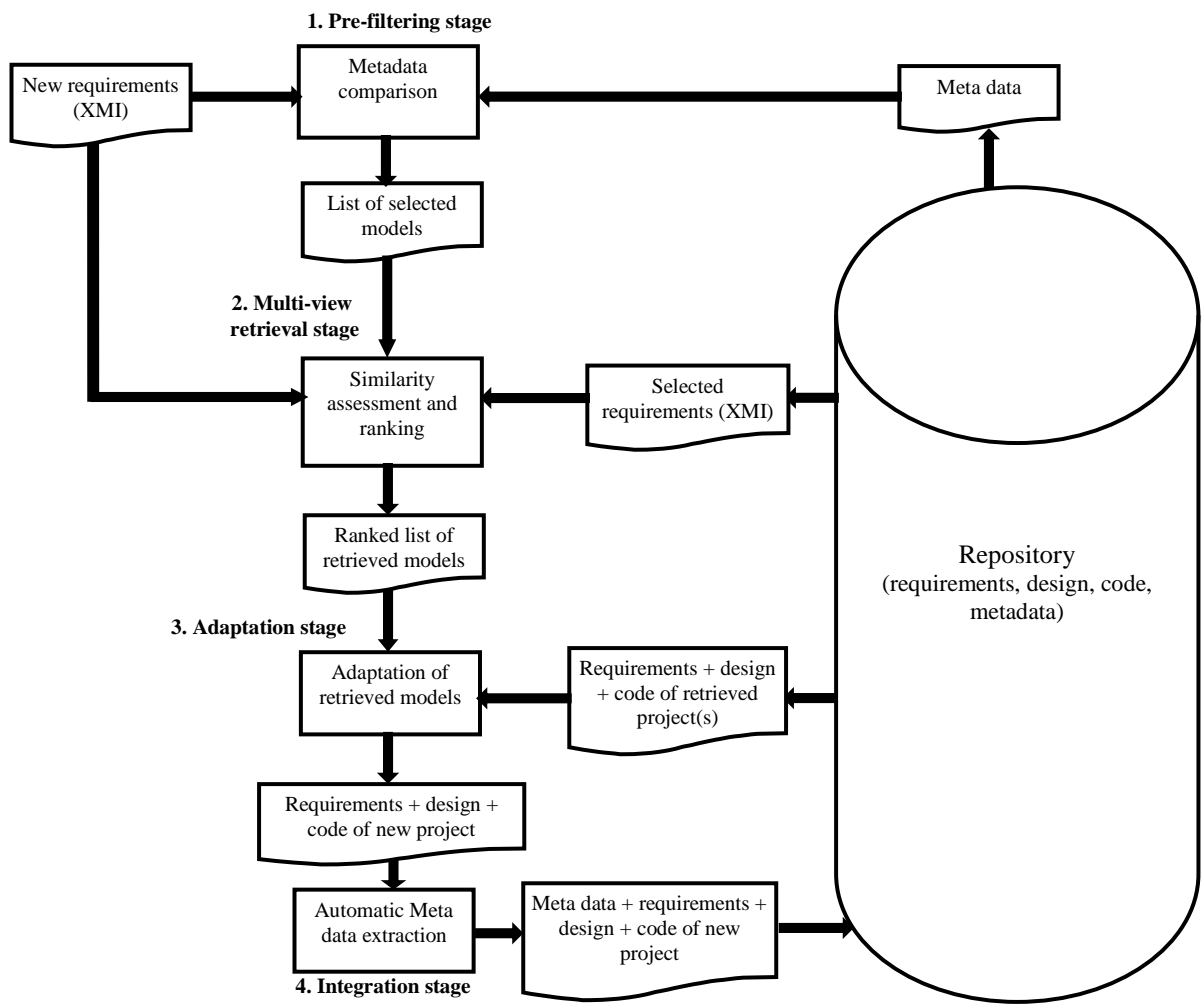

Figure 1.   Schematic diagram of proposed multi-view reuse system

## V. MULTI-VIEW RETRIEVAL USING GENETIC ALGORITHM

System requirements are typically modeled from different but related viewpoints [4]. The division into different views is arbitrary, and often includes at least three views namely structural view, functional view and behavioral view [25-27]. Thus, even though the UML taxonomy of diagrams provides only structure and behavior diagrams, we propose a multi-view retrieval technique that considers three system views: functional view, structural view and behavioral view. Sequence diagrams, class diagrams and state machine diagrams would be considered as representative diagrams of the functional, structural and behavioral views, respectively.

In this section, we describe our retrieval technique which comprises entity matching and similarity scoring. Our initial similarity scoring algorithms are still being fine-tuned, thus we describe only the matching technique which makes use of GA. Furthermore, we describe how our matching technique resolves the different issues raised in Section IV.

Matching refers to mapping an entity in one model to another entity of the same type in the other model to be compared. Once a pair of entities has been mapped, a similarity scoring algorithm can be used to compute their degree of similarity. An entity could be a class, object (i.e. instance of a class) or an entire diagram.

Our proposed approach first maps the classes in the class diagrams of both requirement specifications. A structural similarity measure is then computed from both diagrams. Next, the functional and behavioral similarity scores are computed. Functional similarity is measured by mapping pairs of sequence diagrams in the two models. Similarity between two mapped sequence diagrams is computed using the previously established class mappings, and considering the number and order of messages exchanged between mapped objects in the two sequence diagrams.

State machine diagrams show how an object responds to events according to its current state, and how it enters into new states [24]. Thus, we assume that the previously established class mappings implicitly determine the mapping of the state machine diagrams that depict the behavior of objects of the mapped classes. Finally, multi-view similarity is computed as an aggregate of the three similarity values. Thus, the proposed retrieval method is not expected to produce maximum similarity values for requirement specifications Q and R1 in Fig. 2.

Determining an optimal mapping for the entities in two requirement specifications to be compared is a combinatorial optimization problem. If the entities are arbitrarily mapped, the best possible similarity value may not be obtained. GA can be used to obtain an optimal (or near optimal) mapping of entities in two requirement specifications. GA has been used to solve combinatorial optimization problems such as travelling salesman problem, timetabling and eight queens' chess problem.

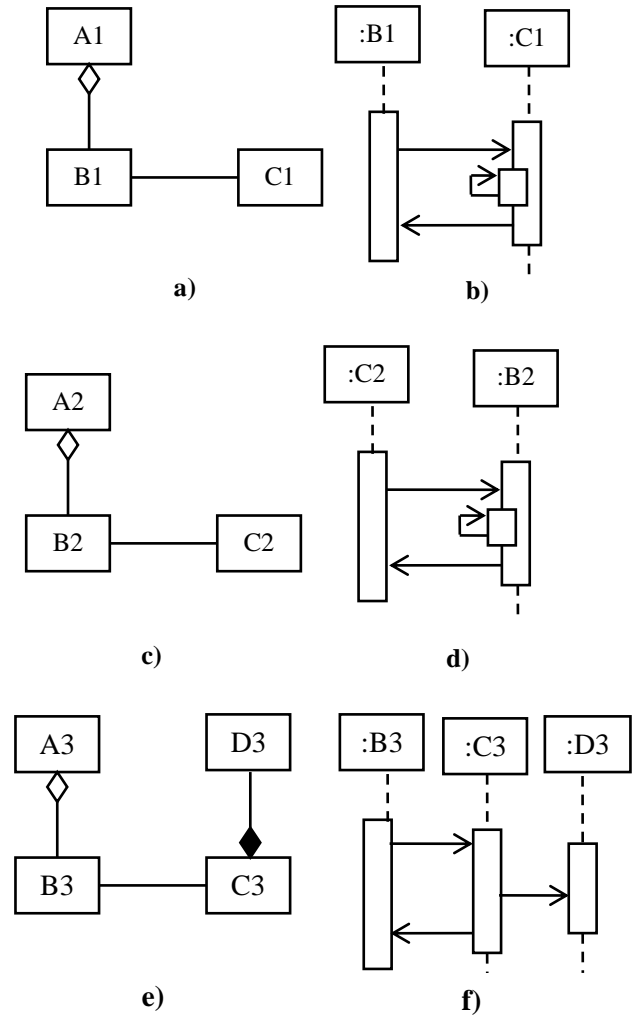

Figure 2. three sample requirement specifications. a) Q's class diagram, b) Q's sequence diagram, c) $R_1$'s class diagram, d) $R_1$'s sequence diagram, e) $R_2$'s class diagram, f) $R_2$'s sequence diagram

Let Q and R be two requirement specifications. Q has cQ classes in its class diagram and sQ sequence diagrams. Similarly, R has cR classes in its class diagram and sR sequence diagrams. Let minC and maxC be the smaller and larger of the values cQ and cR. Likewise, let minS and maxS be the smaller and larger of the values sQ and sR. A suitable encoding of a chromosome to determine a mapping of entities in Q to entities in R is shown in Fig. 3. The chromosome has two parts. The first part of the chromosome handles the mapping of classes in the two class diagrams to be compared, while the second part manages the mapping of sequence diagrams in Q and R. For example, the 2nd, 5th and maxCth class of the requirement having more classes is mapped to the 1st, 2nd and 3rd classes of the requirement having fewer classes. The mapping of sequence diagrams can be inferred similarly from the second part of the chromosome.

Mapping of state machine diagrams in Q and R is not captured in the chromosome because the mapping is implicitly derived from the mapping of classes in the first part of the chromosome. We are currently designing a multi-objective fitness function which will be an aggregate of the structural, functional, and behavioral similarity values. Each of the three similarity values will be computed using appropriate similarity scoring algorithms.

Our proposed retrieval approach properly handles the issue of consistent mapping of classes in class diagrams and sequence diagrams because the classes are first mapped (in the first part of the chromosome), then the established mappings are utilized during the computation of functional similarity values.

As previously explained, consistent mapping of classes in class diagrams and state machine diagrams are implicitly handled. Once a pair of classes has been mapped, the state machine diagrams depicting the behavior of objects of both classes can be compared using a behavioral similarity scoring algorithm.

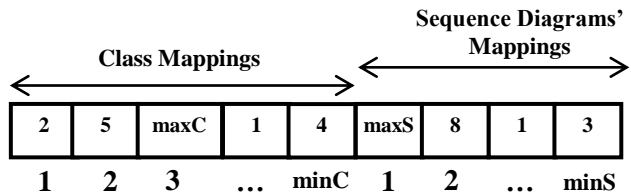

Figure 3. Chromosome design for multi-view retrieval using GA.

The second part of the chromosome handles the mapping of multiple sequence diagrams in two requirement specifications. We have avoided using nested levels of GA similar to the technique in [13] due to the computational complexity of such approach. Instead, we use a multi-objective fitness function. It is also possible to exhaustively compute the similarity scores between pairs of sequence diagrams in the two models then apply a combinatorial optimization algorithm such as Hungarian Algorithm to determine an optimal functional similarity score. We expect this approach to be computationally more expensive than our proposed approach and plan to verify this empirically.

## VI. CONCLUSION AND FUTURE WORK

Software is typically modeled from at least three perspectives: structural view, behavioral view and functional view. In this work, we have classified previous UML reuse work as either multi-view or single-view, depending on whether or not the existing work has considered UML artifacts representing more than one view during retrieval.

We have also raised three issues regarding mapping of entities during multi-view retrieval. We noted that none of the existing multi-view retrieval studies has addressed these issues. A framework for software reuse incorporating multi-view retrieval has been presented. We have shown how GA can be used within the framework to tackle the raised issues.

Our work is currently ongoing hence it is far from being complete. We are currently fine-tuning our similarity scoring algorithms for structural, functional and behavioral views. Furthermore, standard Information Retrieval measures for ranked retrieval such as *Mean Average Precision* and *R Precision* will be used to evaluate our reuse system. In addition, we plan to develop a CASE tool that implements our framework. The CASE tool will accept UML models in XMI format, which is widely supported by prominent CASE tools.

### ACKNOWLEDGMENT

The authors would like to acknowledge the support provided by the Deanship of Scientific Research at King Fahd University of Petroleum & Minerals (KFUPM) under Research Grant 11-INF1633-04.

* Corresponding Author:
Hamza Onoruoiza Salami,
Information and Computer Science Department,
King Fahd University of Petroleum and Minerals, Dhahran, Saudi Arabia,
Email: hosalami@kfupm.edu.sa     Tel:+966-3860-7356